\documentclass[journal]{IEEEtran}
\usepackage{cite}
\usepackage{amsmath,amssymb,amsfonts,amsthm}
\usepackage{algorithm}
\usepackage{algorithmic}
\usepackage{graphicx}
\usepackage{textcomp}
\usepackage{xcolor}
\usepackage{url}
\usepackage{lipsum}
\usepackage{multirow}
\usepackage[normalem]{ulem}
\usepackage{bbding}
\usepackage{pifont}
\usepackage{wasysym}
\usepackage{bbm}
\usepackage{dsfont}
\usepackage{mathtools, nccmath}
\usepackage{makecell}
\usepackage{aligned-overset}
\usepackage{stfloats}
\usepackage{float}
\usepackage{soul}
\usepackage{threeparttable}
\usepackage{empheq}
\ifCLASSOPTIONcompsoc
\usepackage[caption=false, font=normalsize, labelfont=sf, textfont=sf]{subfig}
\else
\usepackage[caption=false, font=footnotesize]{subfig}
\fi
\newfloat{procedure}{tbp}{lop}
\floatname{procedure}{Procedure}

\DeclareMathAlphabet{\mathbcal}{OMS}{cmsy}{b}{n}
\def\BibTeX{{\rm B\kern-.05em{\sc i\kern-.025em b}\kern-.08em
		T\kern-.1667em\lower.7ex\hbox{E}\kern-.125emX}}
\newtheorem{remark}{Remark}

\begin{document}
	
	\title{%
		\huge
		Robust and Secure Near-Field Communication via Curved Caustic Beams
	}
	\author{Shicong~Liu,~\IEEEmembership{Graduate~Student~Member,~IEEE}, Xianghao~Yu,~\IEEEmembership{Senior~Member,~IEEE}, and~Robert~Schober,~\IEEEmembership{Fellow,~IEEE}
		\vspace{-5mm}
		\thanks{
			
			Shicong Liu and Xianghao Yu are with the Department of Electrical Engineering, City University of Hong Kong, Hong Kong (email: sc.liu@my.cityu.edu.hk, alex.yu@cityu.edu.hk).

			Robert Schober is with the Institute for Digital Communications, FAU, 91058 Erlangen, Germany (email: robert.schober@fau.de).
		}
	}
	
	\maketitle
	
\begin{abstract}
	Near-field beamfocusing with extremely large aperture arrays can effectively enhance physical layer security. Nevertheless, even small estimation errors of the eavesdropper's location may cause a pronounced focal shift, resulting in a severe degradation of the secrecy rate. 
	In this letter, we propose a physics-informed robust beamforming strategy that leverages the electromagnetic (EM) caustic effect for near-field physical layer security provisioning, which can be implemented via phase shifts only. 
	Specifically, we partition the transmit array into caustic and focusing subarrays to simultaneously bypass the potential eavesdropping region and illuminate the legitimate user, thereby significantly improving the robustness against the localization error of eavesdroppers. 
	Moreover, by leveraging the connection between the phase gradient and the EM wave departing angle, we derive the corresponding piece-wise closed-form array phase profile for the subarrays. 
	Simulation results demonstrate that the proposed scheme achieves up to an 80\% reduction of the worst-case eavesdropping rate for a localization error of 0.25 m, highlighting its superiority for providing robust and secure communication.
\end{abstract}

\vspace{-2mm}
\section{Introduction}
\bstctlcite{IEEEexample:BSTcontrol}

For sixth-generation (6G) mobile communication systems, extremely large-scale MIMO (XL-MIMO) is expected to become a pivotal technology for achieving unprecedented spectral efficiency and spatial resolution. %
With the growth of the number of elements in XL-MIMO, the associated increasing aperture significantly extends the Rayleigh distance, encompassing more users within the near-field region. {The spherical wavefront in near-field electromagnetic (EM) wave propagation facilitates beamfocusing~\cite{9738442}, which concentrates energy at specific locations, allowing wireless systems to mitigate interference and achieve high data rates~\cite{10504668,10558818}.}

Despite the promising potential of beamfocusing, existing research on near-field secure communication still suffers from several notable limitations. 
First, most studies have assumed perfect channel state information (CSI) of the eavesdropper to be available at the transmitter~\cite{10504668}, which can be overly optimistic in practical passive wiretapping scenarios. 
Second, robust near-field designs often suffer from 
prohibitive computational complexity due to the use of iterative optimization procedures~\cite{8714023}, which hinders their practical application in XL-MIMO systems. 
Furthermore, the CSI uncertainty in the near field is predominantly dictated by \textit{localization} errors, which makes conventional far-field robust optimization techniques, e.g., S-procedure, tailored to tackle norm-bounded \emph{CSI acquisition} errors inapplicable~\cite{11072251}. 
{Although the authors of~\cite{10999042} mitigate CSI uncertainty by injecting artificial noise into the null space of the legitimate channel, such a scheme still relies heavily on the degrees of freedom of the channel, and may suffer from user CSI mismatch. 
} 
Third, the numerous existing robust design approaches still exhibit residual information leakage to eavesdroppers, potentially compromising secure links if the eavesdropper possesses strong detection capabilities. Therefore, developing low-complexity, robust, and nearly absolute secure beamforming strategies for near-field physical layer security remains a critical open challenge. 

To fill this gap, we propose a novel physics-informed robust and secure communication scheme in the near field {by manipulating the envelope of a family of departing EM waves, which is known as the EM caustics}. 
In this work, we design the curved caustic beam trajectory {synthesized by a hardware-efficient passive metasurface} to steer the radiated energy to bypass a circular region defined by the localization error of the eavesdropper. {In this way, information leakage to the designated region is suppressed.} %
To enhance energy efficiency, we introduce a piece-wise trajectory design by partitioning the transmit array into focusing and caustic subarrays. Specifically, array elements with direct paths to the legitimate user that do not intersect the eavesdropping region form a focusing subarray, while the remaining array elements are assigned to a caustic subarray. 
Then, for the caustic subarray, the departing rays are designed to bypass the eavesdropping region, while for the focusing subarray, all departing rays are focused on the legitimate user. 
This approach combines the high energy efficiency of beamfocusing with the security provisioning enabled by EM caustics. 
We finally derive a closed-form phase profile to synthesize the desired curved caustic trajectory, which significantly reduces the computational overhead.

\section{Problem Statement}

\subsection{System Model}
\begin{figure}
	\centering
	\vspace{-2mm}
	\includegraphics[width=0.45\textwidth]{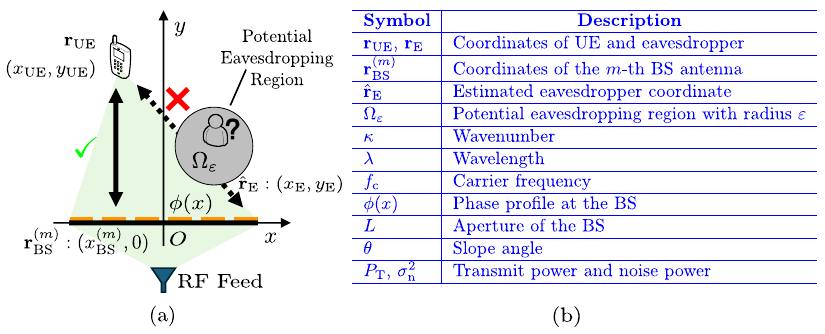}
	\vspace{-3mm}
	\caption{(a) Illustration of near-field secure communication and (b) table with notations. %
	}
	\label{fig:sysmodel}
	\vspace{-3mm}
\end{figure}

Consider a near-field secure communication scenario as illustrated in Fig.~\ref{fig:sysmodel}(a), where a base station (BS) serves a single-antenna legitimate user equipment (UE), and an eavesdropper is wiretapping in the near-field region of the BS array. {The BS is equipped with an $M$-element passive metasurface~\cite{9310290} deployed along the $x$-axis, with the $m$-th array element located at ${\bf r}_{\rm BS}^{(m)} = (x_{\rm BS}^{(m)}, 0)$, and the position of the UE is given by ${\bf r}_{\rm UE} = (x_{\rm UE}, y_{\rm UE})$.} 
{EM waves radiated from the radio frequency (RF) feed port illuminate the metasurface\footnote{The channel between the RF feed and the $m$-th metasurface element can be characterized by a model similar to~\eqref{eq:channelmodel} based on their spatial separations~\cite[Eq.~(6)]{9310290}\cite[Eq.~(1)]{10279595}. As this channel can be calibrated and compensated by the metasurface, further details are thereby omitted for brevity.} and experience phase modulation when passing through the metasurface.}

We adopt the spherical wave model~\cite{10558818} to characterize the near-field wireless channel as 
\begin{equation}
	h\left( {\bf r}_{\rm T},  {\bf r}_{\rm R}\right) = \frac{e^{\jmath \kappa \Vert {\bf r}_{\rm T} - {\bf r}_{\rm R} \Vert}}{\Vert {\bf r}_{\rm T} - {\bf r}_{\rm R} \Vert},
	\label{eq:channelmodel}
\end{equation}
where $\mathbf{r}_\mathrm{T}$ and $\mathbf{r}_\mathrm{R}$ are the coordinates of the transceivers, respectively. In addition, $\kappa = 2\pi/\lambda$ is the wavenumber with $\lambda$ being the carrier wavelength, and $\jmath$ is the imaginary unit such that $\jmath^2 = -1$. 
Therefore, the legitimate channel vector ${\bf h}\in\mathbb{C}^{M\times 1}$ is given by
\begin{equation}
	{\bf h} = \left[ h\left( {\bf r}_{\rm BS}^{(1)},  {\bf r}_{\rm UE}\right),\cdots,h\left( {\bf r}_{\rm BS}^{(M)},  {\bf r}_{\rm UE}\right) \right]^T.
\end{equation}
It can be observed from~\eqref{eq:channelmodel} that near-field channel estimation is in principle a localization problem. In this work, we assume the availability of perfect location information $\{ \mathbf{r}_\mathrm{BS}^{(m)} \}_{m=1}^M$ (and therefore the CSI $\mathbf{h}$) for the BS-UE link via effective near-field localization methods~\cite{10845870}. In contrast, as eavesdroppers typically try to hide their existence from the BS, they are not expected to cooperate with the BS for localization. Hence, the location of the eavesdropper is modeled as $\mathbf{r}_\mathrm{E}=\hat{\mathbf{r}}_\mathrm{E}+\Delta\mathbf{r}$, where $\hat{\mathbf{r}}_\mathrm{E}=(x_\mathrm{E},y_\mathrm{E})$ is the estimate of the eavesdropper's location available at the BS while the localization error is modeled by the term $\Delta\mathbf{r}$. Consequently, the $m$-th element of the eavesdropping channel vector ${\bf h}_{\rm E}\in\mathbb{C}^{M\times 1}$ is modeled as~\cite{9133130}
\begin{equation}
	\begin{aligned}
		\!\!h\left( {\bf r}_{\rm BS}^{(m)},  {\bf r}_{\rm E}\right)&= h\left( {\bf r}_{\rm BS}^{(m)},  \hat{\bf r}_{\rm E} \!+\! \Delta {\bf r}\right),~\Delta {\bf r} \in \Omega_\varepsilon,
		\\
		\Omega_\varepsilon&\triangleq \left\{ {\bf r}=(x,y) \mid \Vert {\bf r} \Vert\leq \varepsilon \right\},
	\end{aligned}
	\label{eq:wiretapping}
\end{equation} 
where $\Omega_\varepsilon$ contains all possible localization errors for the eavesdropper with their norms bounded by $\varepsilon$. Hence, the area depicted by $\Omega_\varepsilon$ is a circle centered at $\hat{\mathbf{r}}_\mathrm{E}$ with radius $\varepsilon$.

\subsection{Problem Formulation}
{In this work, to avoid the high cost and hardware complexity associated with hundreds of phase shifters, we consider employing a passive metasurface to facilitate analog beamforming. Let ${\bf f} = [f_1,\cdots,f_M]^T$ be the phase shift vector of the metasurface at the BS,} 
i.e., $\vert {f}_i\vert = 1,\,\forall 1\leq i\leq M$, and the achievable rates $R_{\rm UE}$ for the legitimate UE and $R_{\rm E}$ for the eavesdropper are respectively given by
\begin{equation}
	R_{\rm UE}\!=\!\log_2\left( 1\!+\!\frac{P_{\rm T}\vert {\bf h}^H {\bf f}\vert^2}{\sigma_{\rm n}^2} \right),~~R_{\rm E} \!=\! \log_2\left( 1\!+\!\frac{P_{\rm T}\vert {\bf h}_{\rm E}^H {\bf f}\vert^2}{\sigma_{\rm n}^2} \right),\notag
\end{equation}
where %
$P_{\rm T}$ denotes the transmit power, and $\sigma_{\rm n}^2$ is the noise power at the receiver. 
The secrecy rate $R_{\rm S}$ between the BS and the legitimate UE is then expressed as
\begin{equation}
	R_{\rm S} = \max\left( 0, R_{\rm UE}-R_{\rm E} \right).\label{eq:rs}
\end{equation}

{Therefore, our design objective is to maximize the worst-case secrecy rate in the potential eavesdropping region by designing the analog beamformer ${\bf f}$ without precise position information about the eavesdropper, which leads to the following optimization problem
\begin{equation}
	\begin{aligned}
		&\underset{{\bf f}}{\rm max} && \underset{\Delta{\bf r}\in \Omega_{\varepsilon}}{\rm min}~{R}_{\rm S}\\[-4pt]
		&\mathrm{s.t.}&&\vert {\bf f} [m]\vert = {1}/{\sqrt{M}},\quad \forall~1\leq m\leq M.%
	\end{aligned}
	\label{eq:P1}
\end{equation}}

\section{Proposed Beam Trajectory Design}

{Instead of directly solving the robust and secure communication problem~\eqref{eq:P1} via relaxation and iterative numerical optimization, in this section, we propose a physics-informed method that leverages the geometric information characterized by $\Omega_\varepsilon$ and curved caustic beam trajectories. 
Specifically, 
we develop a novel caustic beamforming scheme by partitioning the transmit array into caustic and focusing subarrays, where the caustic subarray suppresses the eavesdropping rate $R_{\rm E}$, while the focusing subarray enhances the legitimate rate $R_{\rm UE}$ to jointly improve the worst-case secrecy rate in~\eqref{eq:P1}.}

\subsection{Phase Gradients}
{We first present how the angles of departure (AoDs) of the caustic beam are manipulated by the phase gradient.} {To facilitate a tractable derivation, we assume continuous phase variation on the BS array, which will be discretized in the subsequent numerical simulations for performance evaluation.} 
Without loss of generality, we consider a monochromatic EM wave $U(x,y)$ with wavelength $\lambda$ in the $xOy$ plane satisfying the Helmholtz equation in a homogeneous and isotropic medium as
\begin{equation}
	\left( \nabla^2 + \kappa^2 \right) U(x,y) = 0,\label{eq:helmholtz}
\end{equation}
where $\nabla^2$ is the Laplace operator. 
Let the trial solution be
\begin{equation}
	U(x,y) = A(x,y)e^{\jmath\kappa D(x,y)},\label{eq:trialsol}
\end{equation}
where $A(x,y)$ and $D(x,y)$ are arbitrary real amplitude- and distance-related functions, respectively. Substituting~\eqref{eq:trialsol} into~\eqref{eq:helmholtz}, we have
\begin{equation}
	\nabla^2 A + 2\jmath \kappa \nabla A\cdot \nabla D + \jmath\kappa A\nabla^2 D-A\kappa^2\vert \nabla D \vert^2 + \kappa^2 A = 0,\label{eq:temp1}
\end{equation}
where operator $\cdot$ denotes the inner product between vectors, and we omit the dependence on variables $x$ and $y$ for brevity. Then, dividing~\eqref{eq:temp1} by $\kappa^2$, we further have%
\begin{equation}
	\frac{\nabla^2 A}{\kappa^2} + \frac{2\jmath \nabla A\cdot \nabla D + \jmath A\nabla^2 D}{\kappa} + A\left(1 - \vert \nabla D \vert^2\right) = 0.\label{eq:reorg}
\end{equation}
Since wavenumber $\kappa$ is typically large in wireless communications, 
the impact of $\nabla^2 A$ becomes insignificant in~\eqref{eq:reorg} due to the scaling with $1/\kappa^2$. Besides, for a slow-varying amplitude function $A$, the Laplacian $\nabla^2 A$ is also generally small. Therefore, we can safely ignore this term and obtain a surrogate equation of~\eqref{eq:reorg} as
\begin{equation}
	\frac{\jmath}{\kappa}\left( 2\nabla A\cdot\nabla D + A\nabla^2 D \right) + A\left(1 - \vert \nabla D \vert^2\right) = 0,\label{eq:surrogate}
\end{equation}
where the real and imaginary parts equal zero simultaneously, yielding
\begin{equation}
	\Vert \nabla D (x,y) \Vert = \left\Vert \left( \frac{\partial D(x,y)}{\partial x}, \frac{\partial D(x,y)}{\partial y} \right) \right\Vert=1\label{eq:eikonal}
\end{equation}
and
\begin{equation}
	2\nabla A(x,y)\cdot\nabla D(x,y) + A(x,y)\nabla^2 D(x,y) = 0.
\end{equation}
For an EM wave $U(x,y)$, the level surface of the distance term $D(x,y)$ is known as the wavefront, and the gradient $\nabla D(x,y)$ is always perpendicular to the wavefront, which indicates the propagation direction. Since the magnitude of this gradient is equal to a constant according to~\eqref{eq:eikonal}, it follows that, if we can design and control the gradient components {on the transmitting interface (e.g., the $x$-component on the $y=0$ surface)}, we can then control the propagation direction of the EM wave. 

In other words, given a desired AoD $\theta$, see Fig.~\ref{fig:caustic}(a), one straightforward design of 
the phase profile of $U(x,y)$ in~\eqref{eq:trialsol}, i.e., $\phi(x,y) = \kappa D(x,y)$, is given by%
\begin{equation}
	\left.\frac{\partial \phi(x,y)}{\partial x}\right\vert_{y=0} =\kappa_x =\kappa \cos \theta. \label{eq:eikonal_x}
\end{equation}
In this case, according to~\eqref{eq:eikonal}, the $y$-component of the phase gradient has to be $[{\partial \phi(x,y)}/{\partial y}]_{y=0} =\kappa_y =\kappa \sin \theta$. Consequently, the departing beam propagates along the direction ${\boldsymbol{\kappa}} = (\kappa_x,\kappa_y) =  (\kappa\cos\theta,\kappa\sin\theta)$. 

\begin{remark}
	Given specific boundary conditions at an interface, i.e., $y=0$, the solution in~\eqref{eq:eikonal_x} is referred to as the generalized Snell's law~\cite{generalizedsnell}. Such a characteristic can also guide the design of beam steering and focusing by solving~\eqref{eq:eikonal_x} with specific variations of the AoD $\theta$. %
	
	For instance, the conventional beam steering scheme along a fixed AoD $\theta$, as shown in Fig.~\ref{fig:caustic}(a), can be regarded as the solution to~\eqref{eq:eikonal_x} 
	as $\phi_{\rm Steering}(x) = \kappa\cos\theta \cdot x + C$ with $C$ being an arbitrary constant. In this case, the phase profile is a linear function of $x$ as shown in Fig.~\ref{fig:caustic}(d). 
	
	Similarly, to focus all the transmit energy on a UE located at ${\bf r}_{\rm UE}= (x_{\rm UE}, y_{\rm UE})$, the AoD $\theta$ has to vary with $x$ as
	\begin{equation}
		\cos\theta = {\left( x_{\rm UE}-x \right)}/{\sqrt{(x_{\rm UE}-x)^2+y^2_{\rm UE}}},\label{eq:focusingtheta}
	\end{equation}
	as shown in Fig.~\ref{fig:caustic}(b). The corresponding focusing phase profile is then calculated by substituting~\eqref{eq:focusingtheta} into~\eqref{eq:eikonal_x} as
	\begin{align}
		\phi_{\rm Foc}(x) = \int \kappa\cos\theta\,{\rm d}x=-\kappa\Vert {\bf r}_{\rm UE} - (x,0) \Vert + C,\label{eq:foc}
	\end{align}
	which aligns with conventional beamfocusing~\cite{9738442}, as illustrated in Fig.~\ref{fig:caustic}(e).
\end{remark}

\subsection{Caustic Trajectory Design}
\begin{figure}[t]
	\centering
	\includegraphics[width=0.425\textwidth]{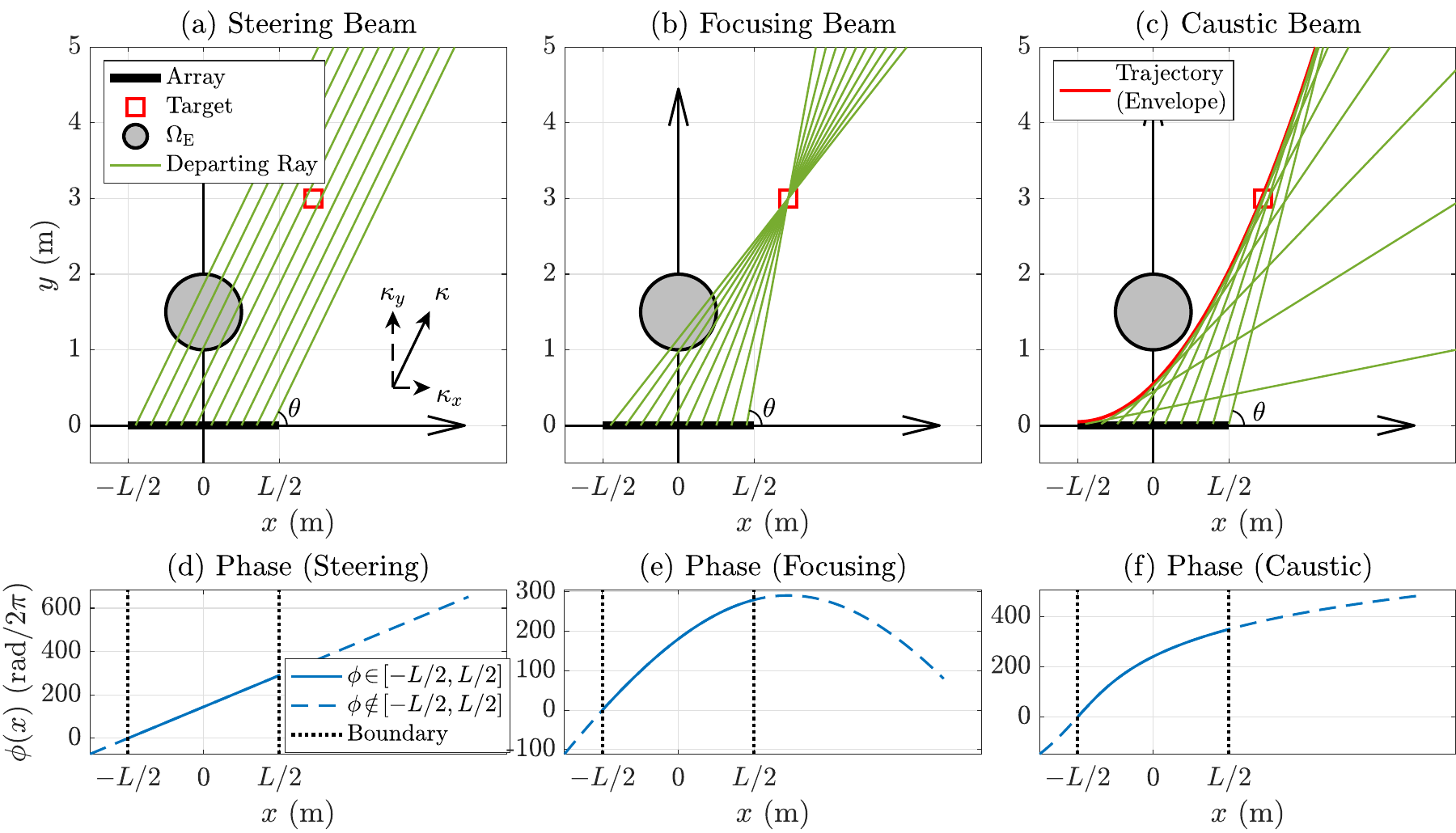}
	\vspace{-3mm}
	\caption{Illustration of (a) steering beam, (b) focusing beam, and (c) caustic beam, as well as the corresponding phase profiles (d), (e), and (f).}\label{fig:caustic}
	\vspace{-3mm}
\end{figure}

Based on the described EM propagation characteristics, we can not only achieve beam steering and focusing, but also tailor the curved caustic beams. {A caustic is the envelope of a family of departing EM waves, stretching a focal point into a continuous trajectory.} By elongating the focal point, the curved trajectory is able to bypass the undesired eavesdropping region $\Omega_\varepsilon$ to support nearly absolute secure communication.

In particular, we choose a spatial trajectory $\mathcal{T}$ that detours around the area $\Omega_{\varepsilon}$, and enforce that all beams radiated by the array are tangents to this trajectory, as shown in Fig.~\ref{fig:caustic}(c). In this way, the departing rays form an envelope that coincides with the prescribed trajectory, ensuring no departing EM waves pass through the area $\Omega_{\varepsilon}$.

For instance, consider a curved spatial trajectory $\mathcal{T}\triangleq\left\{ (x,y) \mid y = f(x)\right\} \subset \mathbb{R}^2$ 
and an arbitrary point $(\xi,f(\xi))$ on it. Then, the tangent line at $(\xi,f(\xi))$ intersects the array at $(x_\xi,0)$, where the $x$-intercept is given by
\begin{equation}
	x_\xi = \xi -\frac{f(\xi)}{f^\prime(\xi)},\quad x\in\left[-\frac{L}{2},\frac{L}{2}\right],
\end{equation}
and $L = (M-1)d$ is the aperture of the BS array, with $d$ being the element spacing. Since the slope of the tangent line is given by the first order derivative $f^\prime(\xi) = \tan\theta$,~\eqref{eq:eikonal_x} then yields
\begin{equation}
	\left.\frac{\partial \phi(x)}{\partial x}\right\vert_{x=x_\xi} =\kappa\cos\theta = \frac{\kappa}{\sqrt{1\!+\!\left( f^\prime(\xi) \right)^2}}.\label{eq:eikonal_var}
\end{equation}
The desired phase profile can then be obtained by solving~\eqref{eq:eikonal_var}.

It is also worth noting that, for trajectories characterized by basic elementary functions, the phase profile solutions corresponding to~\eqref{eq:eikonal_var} are well established~\cite{Froehly:11}. For example, the solution to~\eqref{eq:eikonal_var} with the quadratic trajectory $f(x) = (x/a)^2$ shown in Fig.~\ref{fig:caustic}(f) is given by
\begin{equation}
	\phi_{\rm Quad}(x) = \frac{\kappa a^2}{4}{\rm asinh}\left( \frac{4x}{a^2} \right).
\end{equation}
In addition, when $f$ is a parabolic function with acceleration parallel to the array axis, the beam synthesized by~\eqref{eq:eikonal_var} reduces to the well-known Airy beam~\cite{guerboukha_curving_2024}.

\subsection{Proposed Secure Beam Trajectory}

\begin{figure}[t]
	\centering
	\vspace{-3mm}
	\includegraphics[width=0.225\textwidth]{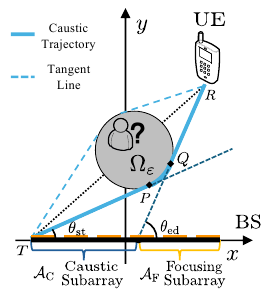}
	\vspace{-3mm}
	\caption{The proposed piece-wise trajectory design.}\label{fig:proposed}
	\vspace{-3mm}
\end{figure}

In this subsection, we propose a practical and generally applicable piece-wise caustic trajectory design, which improves the energy efficiency by integrating caustic shaping with beamfocusing in the near field. The corresponding phase profile $\phi(x)$ is derived in closed form. {Note that the derivations in this subsection rely on the spatial geometry shown in Fig.~\ref{fig:proposed}, while they can be readily generalized to scenarios with different geometric configurations of UE and eavesdroppers.}

First, according to the availability of direct line-of-sight (LoS) paths between BS array and the UE\footnote{In this letter, we focus on the single-user case to reveal the underlying physical mechanism and obtain a more tractable closed-form phase profile. However, the proposed scheme can be extended to scenarios with multiple users/eavesdroppers by partitioning the entire BS array into multiple caustic subarrays to bypass multiple potential eavesdropping regions and multiple focusing subarrays for user-specific secure beam design.}, we 
partition the BS array region $\mathcal{A} = \{ x\mid -L/2\leq x\leq L/2 \}$ into a caustic subarray $\mathcal{A}_{\rm C}$ (without LoS paths) and a focusing subarray $\mathcal{A}_{\rm F}$ (with LoS paths), as illustrated in Fig.~\ref{fig:proposed}. The caustic subarray shapes its departing rays to bypass the eavesdropping region $\Omega_{\varepsilon}$, while the focusing subarray concentrates departing rays on the UE. 

Then, we can draw 
two tangent lines from the far end of the BS array $T = (-L/2,0)$ and the UE $R = {\bf r}_{\rm UE}$ to the circular region $\Omega_{\varepsilon}$, and select the two tangent segments $\overline{TP}$ and $\overline{QR}$ that lie closer to the focusing subarray. The corresponding tangent points $P$ and $Q$ then define the minor arc $\widehat{PQ}$ on the circle. Finally, 
a piece-wise caustic trajectory is constructed by %
\begin{equation}
	\mathcal{T} = \overline{TP}\cup \widehat{PQ}\cup \overline{QR},\label{eq:caustictrajectory}
\end{equation}
as depicted in Fig.~\ref{fig:proposed}. In the following, according to the constructed trajectory $\mathcal{T}$, we derive the phase profiles for the caustic and focusing subarrays, respectively.

\subsubsection{Caustic Subarray}
The tangent line with slope angle $\theta$ satisfies
\begin{equation}
	y = \tan\theta \left( x- x_t\right) + y_t,\label{eq:tangentline}
\end{equation}
where $(x_t,y_t)$ on the boundary circle of $\Omega_{\varepsilon}$ is defined by
\begin{equation}
	x_t = x_{\rm E} + \varepsilon \sin\theta,\quad
	y_t = y_{\rm E} - \varepsilon \cos\theta.
	\label{eq:tangentpoint}
\end{equation}
Substituting~\eqref{eq:tangentpoint} into~\eqref{eq:tangentline}, we have the $x$-intercept 
on the BS array as
\begin{equation}
	x_\theta = -\frac{y_{\rm E} - \varepsilon\cos\theta}{\tan\theta}+x_{\rm E}+\varepsilon\sin\theta.
	\label{eq:xt}
\end{equation}
Since $\theta$ in~\eqref{eq:xt} cannot be represented by a function of $x$ in a simple form, directly solving $\phi(x)$ can be difficult. Hence, we consider solving $\phi(\theta)$ first. According to~\eqref{eq:xt}, we have
\begin{equation}
	\frac{\partial x_\theta}{\partial \theta} = \frac{y_{\rm E} - \varepsilon \cos\theta}{\sin^2\theta}.\label{eq:partialxt}
\end{equation}
Substituting~\eqref{eq:partialxt} into~\eqref{eq:eikonal_var}, we have
\begin{equation}
	\frac{\partial \phi(\theta)}{\partial \theta} = \frac{\partial \phi(\theta)}{\partial x_\theta} \frac{\partial x_\theta}{\partial \theta}  =\frac{y_{\rm E} - \varepsilon \cos\theta}{\sin^2\theta} \cdot \kappa\cos\theta,
\end{equation}
which further yields
\begin{equation}
	\phi(\theta) = \kappa\left( \varepsilon \theta +  \frac{\varepsilon}{\tan\theta} - \frac{y_{\rm E}}{\sin\theta}\right),\quad\theta_{\rm st}\leq \theta\leq \theta_{\rm ed},
\end{equation}
with $\theta_{\rm st}$ and $\theta_{\rm ed}$ being the slope angles of the tangent line segments $\overline{TP}$ and $\overline{QR}$, respectively, as marked in Fig.~\ref{fig:proposed}. {The constant bias of the integral is omitted for brevity.} According to the relationship between $x$ and $\theta$ in~\eqref{eq:xt}, the closed-form expression for $\phi(x)$ is then given by
\begin{equation}
	\phi(x)=\kappa\left( 2\varepsilon{\rm atan}\left( \frac{x-x_{\rm E}+S(x)}{\varepsilon+y_{\rm E}} \right)-S(x) \right),\label{eq:phix}
\end{equation}
where
\begin{equation}
	S(x) = \sqrt{\left( x-x_{\rm E} \right)^2+y_{\rm E}^2 -\varepsilon^2}.
\end{equation}
\subsubsection{Focusing Subarray}
Since the focusing subarray has direct LoS paths to the UE, we directly concentrate all the transmit energy on the UE to boost the legitimate rate $R_\mathrm{UE}$. The phase profile for the focusing subarray is given by~\eqref{eq:foc}.

In summary, the overall phase profile of the proposed trajectory~\eqref{eq:caustictrajectory} is given by
\begin{equation}
	\phi(x) =
	\begin{cases}
		\eqref{eq:phix},& x\in \mathcal{A}_{\rm C}\\
		-\kappa\sqrt{(x_{\rm UE}-x)^2+y^2_{\rm UE}} + C,&x \in \mathcal{A}_{\rm F}
	\end{cases},\label{eq:overallsol}
\end{equation}
where $C$ is a constant to ensure the continuity of $\phi(x)$.
\section{Main Results}
\subsection{Simulation Setup}
In the simulation, the carrier frequency is set as $f_{\rm c} = 28$ GHz. The BS array is equipped with $M = 256$ array elements with half-wavelength spacing $d = \lambda/2$. The transmit power of the BS is set as $P_{\rm T} = 20$ dBm, and the noise power at the receiver is $\sigma_{\rm n}^2 = -50$ dBm. The continuous phase solution in~\eqref{eq:overallsol} is uniformly sampled and applied to the discrete metasurface. 
The following benchmarks are adopted to show the effectiveness of the proposed method.
\begin{itemize}
	\item {\bf Optimal Secure Focusing}~\cite{5485016}: Assuming perfect CSI, the beamforming vector is chosen as the dominant generalized eigenvector of the matrix pencil $({\bf I}_M + \frac{P_{\rm T}}{\sigma_{\rm n}^2}{\bf h}{\bf h}^H,\ {\bf I}_M + \frac{P_{\rm T}}{\sigma_{\rm n}^2}{\bf h}_{\rm E}{\bf h}_{\rm E}^H)$, where ${\bf I}_M$ denotes the identity matrix of order $M$.
		{\item {\bf Norm-Bounded Design}~\cite{11072251}: The worst-case secrecy rate in the eavesdropping region $\Omega_{\varepsilon}$ is maximized by the S-procedure via relaxation of the distance constraint to a channel norm error.}
	\item {{\bf ADMM}~\cite{8714023}: Problem~\eqref{eq:P1} is relaxed and solved iteratively by a Dinkelbach approach.}
	\item {\bf Proposed Caustic Beam}: The piece-wise phase profile is designed according to~\eqref{eq:overallsol}.
	\vspace{-3mm}
\end{itemize}
\subsection{Caustic Beam Visualization}
\begin{figure}[t]
	\centering
	\vspace{-3mm}
	\includegraphics[width=0.4\textwidth]{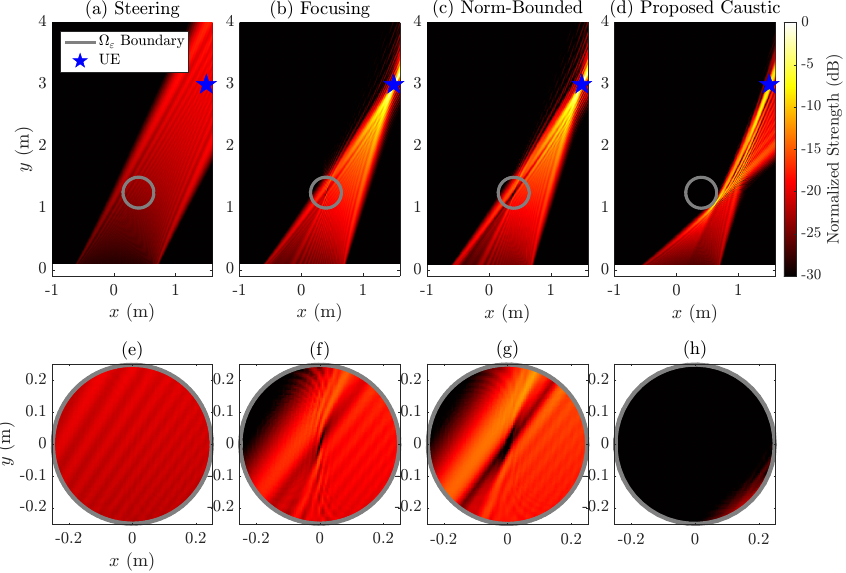}
	\vspace{-3mm}
	\caption{The visualization of spatial beams of different phase profiles, with corresponding radiation details in the region $\Omega_{\varepsilon}$ zoomed in. %
	}\label{fig:four}
	\vspace{-3mm}
\end{figure}

To provide an intuitive and clear illustration, we first visualize the resulting spatial beams\footnote{We consider a pessimistic case with the eavesdropper located close to the BS, which can be a typical scenario in indoor environments~\cite{guerboukha_curving_2024}}. %
The circular eavesdropping region $\Omega_{\varepsilon}$ is centered at $\hat{\mathbf{r}}_\mathrm{E}=(0.4,1.25)$ m with radius $\varepsilon = 0.25$ m, and the UE is located at $(1.5,3)$ m. 
Fig.~\ref{fig:four} depicts the normalized radiation strength achieved by different phase profiles, where the eavesdropping region $\Omega_{\varepsilon}$ is zoomed in for a better view. 
The beam steering scheme in Fig.~\ref{fig:four}(a) radiates energy rather uniformly onto the UE direction $\theta = \arctan(y_{\rm UE}/x_{\rm UE})$, causing high leakage across the entire region $\Omega_{\varepsilon}$. In addition, the baseline schemes in Figs.~\ref{fig:four}(b) and (c) 
can only suppress the leakage in a very limited region within $\Omega_{\varepsilon}$, as depicted in Figs.~\ref{fig:four}(f) and~\ref{fig:four}(g). 

In contrast, the proposed method shown in Fig.~\ref{fig:four}(d) successfully steers the energy to bypass the eavesdropping region, which guarantees nearly absolute secure communication. 
Note that residual leakage still persists at the edge of the eavesdropping region, as shown in Fig.~\ref{fig:four}(h), which is attributed to the omission of the term $\nabla^2 A/\kappa^2$ in~\eqref{eq:reorg} for a tractable analysis. {This indicates that underestimating the radius $\varepsilon$ may degrade the worst-case secrecy rate performance by introducing more residual leakage. To avoid this, we can deliberately increase the radius $\varepsilon$ of the eavesdropping region by a small margin to ensure absolutely secure communication.}

It is also noteworthy that the schemes in Figs.~\ref{fig:four}(b) and~\ref{fig:four}(c), which employ non-unit-modulus beamforming weights, cannot be realized using phase shifts only. Despite this relaxation, they fail to match the performance of our proposed method, primarily because we exploit physics-informed EM properties as domain knowledge to bypass the eavesdropping region.

\vspace{-2mm}

\subsection{Secrecy Rate}

\begin{figure}
	\centering
	\includegraphics[width=0.4\textwidth]{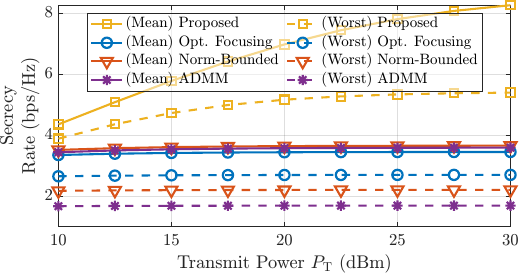}
	\vspace{-3mm}
	\caption{Mean and worst-case rate performance versus transmit power $P_{\rm T}$.}%
	\vspace{-3mm}
	\label{fig:mean_worst}
\end{figure}

We further investigate the secrecy rate performance with varying transmit power $P_{\rm T}\in[10,30]$ dBm. 
As shown in Fig.~\ref{fig:mean_worst}, the proposed scheme consistently achieves superior secrecy rate performance in both mean and worst-case secrecy rate scenarios, {indicating superior robustness over other comparing schemes}. This is because of the significant suppression of energy leakage into the potential eavesdropping region, which leads to a substantial decrease of $R_{\rm E}$ in~\eqref{eq:rs}.
In the high SNR near-field region, additional transmit power leads to an increase in both legitimate and eavesdropping rates, and therefore, the secrecy rate tends to remain constant. In contrast, the proposed curved caustic beams can always bypass the potential eavesdropping region. Hence, more transmit power illuminating the UE leads to a rapid increase in secrecy rate.%

\vspace{-2mm}

\subsection{Computational Complexity}

\begin{table}[t]
	\centering
	\caption{Computational Complexity and Mean Execution Time}\label{table:1}
	\begin{threeparttable}
		\begin{tabular}{c|c|cc}
			\hline\hline
			\multirow{2}{*}{\textbf{Methods}} & \multirow{2}{*}{\begin{tabular}[c]{@{}c@{}}\!\!\textbf{Computational}\!\!\\ \textbf{Complexity}  \end{tabular}} & \multicolumn{2}{c}{\bf Execution Time (s)} \\ \cline{3-4} %
			&                                               & $M=64$         & $M=256$         \\ \hline
			\!\!Opt. Focusing~\cite{5485016}\!\!&        \!\!$\mathcal{O}(3M^3+M^2)$\!\!                                                                                 &   $2.91\!\times\!10^{-3}$            &   \!\!$1.01\!\times\!10^{-1}$\!\!              \\ \hline
			\!\!{ Norm-Bounded}~\cite{11072251}\!\!&         $\sim\mathcal{O}(M^{4.5})$                                                                                     &   \!\!$1.81$\!\!             &   \!\!$27.4$\!\!             \\ \hline
			\!\!ADMM~\cite{8714023}\!\!&         $\sim\mathcal{O}(M^{3})$                                                                                     & \!\!$3.05\times 10^{-2}$\!\!             & \!\!$1.04\times 10^{-1}$\!\!             \\ \hline
			\!\!Proposed\!\!&       N/A                                                                                     &     $7.61\!\times\!10^{-4}$           &   $9.64\!\times\!10^{-4}$              \\ \hline\hline
		\end{tabular}
	\end{threeparttable}
	\vspace{-3mm}
\end{table}

We finally evaluate the computational complexity of the proposed scheme. The computational complexity and the mean execution time are shown in Table~\ref{table:1}, where the complexity is evaluated by the number of multiplications, and the time is averaged over $1,000$ realizations. The proposed scheme shows a significantly shorter execution time than the baseline methods. Moreover, the execution time of the proposed scheme increases only marginally by about $25\%$ as $M$ increases by four times, which shows its scalability in XL-MIMO systems.

\section{Conclusion}

In this letter, we exploited EM wave caustics to facilitate secure near-field communication in the presence of eavesdropper localization errors. 
We first designed a curved caustic trajectory that bypasses the eavesdropping region, and then derived the corresponding phase profile to generate the desired caustic beam. Numerical results demonstrated substantial gains in both the mean and worst-case secrecy rate performance, highlighting the potential of the proposed caustic scheme for providing robust physical layer security. {In practice, hardware imperfections, such as phase quantization and phase noise, may cause distortion of the synthesized trajectory. Thus, additional studies are needed to further evaluate and improve practical deployability.
}

\bibliographystyle{IEEEtran}
\bibliography{IEEEabrv,references}

\end{document}